\documentclass[12pt]{iopart}
\usepackage{latexsym}
\usepackage{graphicx}
\usepackage{epsfig}
\usepackage{amsfonts}
\usepackage{amssymb}

\begin{document}

\title{Sideband cooling and coherent dynamics in a microchip multi-segmented ion trap}

\author{Stephan Schulz \footnote{corresponding author:
stephan.schulz@uni-ulm.de}, Ulrich Poschinger, Frank Ziesel, and Ferdinand
Schmidt-Kaler}

\address{Universit\"at Ulm, Institut f\"ur Quanteninformationsverarbeitung,
Albert-Einstein-Allee 11, D-89069 Ulm, Germany}

\begin{abstract}
Miniaturized ion trap arrays with many trap segments present a promising architecture
for scalable quantum information processing. The miniaturization of segmented linear
Paul traps allows partitioning the microtrap in different storage and processing zones. The
individual position control of many ions  - each of them carrying qubit
information in its long-lived electronic levels -  by the external trap control voltages is
important for the implementation of next generation large-scale quantum algorithms.

We present a novel scalable microchip multi-segmented ion trap with two different
adjacent zones, one for the storage and another dedicated for the processing of quantum
information using single ions and linear ion crystals: A pair of radio-frequency driven
electrodes and 62 independently controlled DC electrodes allows shuttling of single
ions or linear ion crystals with numerically designed axial potentials at axial and
radial trap frequencies of a few MHz. We characterize and optimize the microtrap using
sideband spectroscopy on the narrow $S_{1/2} \leftrightarrow D_{5/2}$ qubit transition
of the $^{40}$Ca$^+$ ion, demonstrate coherent single qubit Rabi rotations and optical
cooling methods. We determine the heating rate using sideband cooling measurements to
the vibrational ground state which is necessary for subsequent two-qubit quantum logic
operations.  The applicability for scalable quantum information processing is proven.

\end{abstract}

\pacs{37.10.Ty, 37.10.-x, 32.80.Qk, 03.67.Lx}


\newpage

\tableofcontents

\newpage

\section{Introduction}
Long term goal of our research is the large-scale quantum computer
\cite{CHUANG,ROADMAPS} with multiple ions and linear ion crystals for encoding complex
entangled states, with optimized laser pulses for processing quantum information
(QIPC), and with well suited time-dependent trap control voltages to shuttle quantum
information between a central processing unit and quantum memories or qubit read-out
sections. On the way towards this goal we present in this paper the most complex microstructured
ion trapping device with a large number of control segments such that in future order of
10$^2$ qubits might be processed. Currently we do not exploit at all the complexity of
our segmented trapping device and the corresponding options for large-scale quantum
processing. The purpose of this paper, however, is describing the trap device,
characterizing its basic performance and proving necessary building blocks for quantum
gates with ions.

The work presented here should be seen in context with pioneering work to fabricate
specialized ion microtraps for QIPC (instead of using mm-sized conventional designs
\cite{NEUHAUSER1978,ROOS1999,ROOSPHD,SCHMIDT2003}) by using microscale planar
\cite{SEIDLIN2006,BROWN2007}, linear three-dimensional
\cite{STICK2006,LEIBFRIED2004,TURCHETTE2000} or even more complex geometries
\cite{HENSINGER2006}. Gold plated substrates allow electrode structures with gaps of a
few $\mu$m, but semiconductor structuring may allow in future even finer and more
complex traps \cite{BROWNUTT2006}. Shuttling protocols for single ions
\cite{ROWE2002,REICHLE2006a,HUCUL2007,HUBER2007} have been tested and even used for
implementing quantum algorithms \cite{BARRETT2004,REICHLE2006b}. On the other hand, the
application of well controlled laser pulses has lead to the entanglement of up to eight
ions \cite{HAEFFNER2005} and the realization of complex quantum algorithms like
e.g. teleportation \cite{RIEBE2004}.

The paper is organized as follows: First we outline the steps for the fabrication of
the multi-segmented linear ion trapping device and the integration in the overall
experimental setup. Cold crystals of trapped $^{40}$Ca$^+$ ions are showing the
elementary operation of this device. For refined studies of trapped ions and the
properties of the trap we employ quantum jump spectroscopy on the
narrow $S_{1/2} \leftrightarrow D_{5/2}$ quadrupole
transition, where all components of the ion motion are resolved easily. Investigating
the micromotional sideband allows shifting of the ion to the center of the
radio-frequency (RF) trapping field. Since we are observing a very small variation of
the necessary compensation voltage only we conclude that charging effects are indeed
effectively impeded by the advanced trap construction. If the ion is excited by a
narrow band laser source, we observe Rabi flopping as we vary the pulse duration.
Finally, we study the secular sidebands and sideband cool a single ion from Doppler
temperature down to the vibrational ground state. Cooling time and trap heating time
are revealed. Finally, in section 5, we sketch future applications.

\newpage

\section{Microchip ion trap}
\begin{figure}[htp]
\resizebox{0.99\hsize}{!}{\includegraphics*{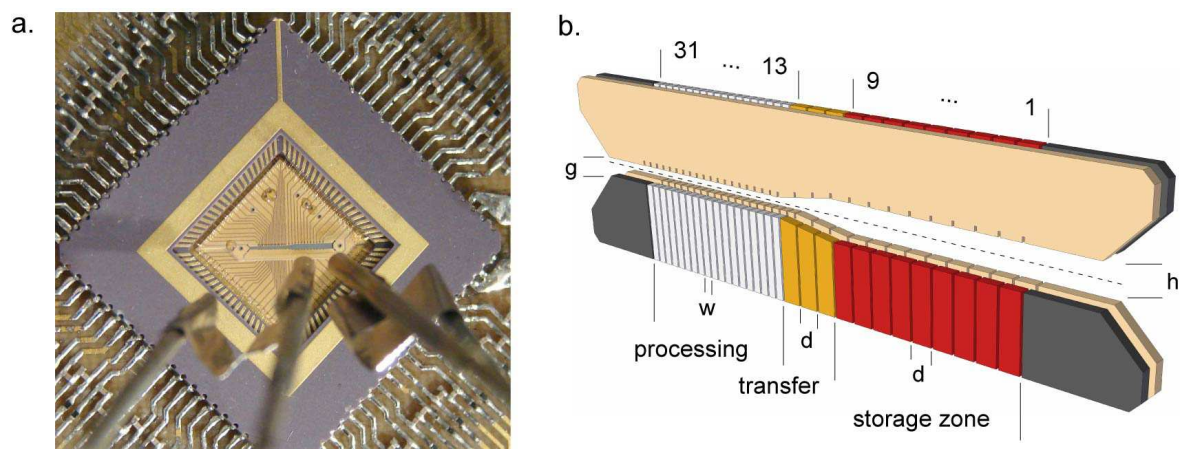}} \caption{(a) Ulm microchip trap
installed in the vacuum housing and connected to the filter board. In the front are the
two calcium ovens. (b) Scheme of the trap: 31 electrode pairs of the different zones for storage,
transfer and processing are characterized by the slit width g and h and the electrode dimensions w and d.} \label{design}
\end{figure}

The Ulm microchip trap is a two-layer electrode design modeled as logical continuation
of conventional linear Paul traps (figure~\ref{design}a). The storage zone is connected
to the processing zone
by a multi-segment transfer region in order to achieve a smooth passage avoiding
vibrational excitation (figure~\ref{design}b). The electrode layers are as symmetric as
possible for well-balanced electric potentials on the trap axis such that
micromotion should be minimized.
Thus, the non-segmented RF segments are cut with notches precisely opposite to
the separation cuts in the DC electrodes to improve the axial electric
field symmetry. The microtrap is assembled in a UHV compatible ceramic chip carrier
with the advantage of an exceptional alignment precision of the electrode layers.
Furthermore easy electric connectivity, exchange usability and in the near future the integration in UHV
compatible electronic boards with fast digital-analog converters and digital
RF synthesizers prioritize this proof of concept.

\subsection{Design and fabrication}

\begin{figure}[htp]
\resizebox{0.99\hsize}{!}{\includegraphics*{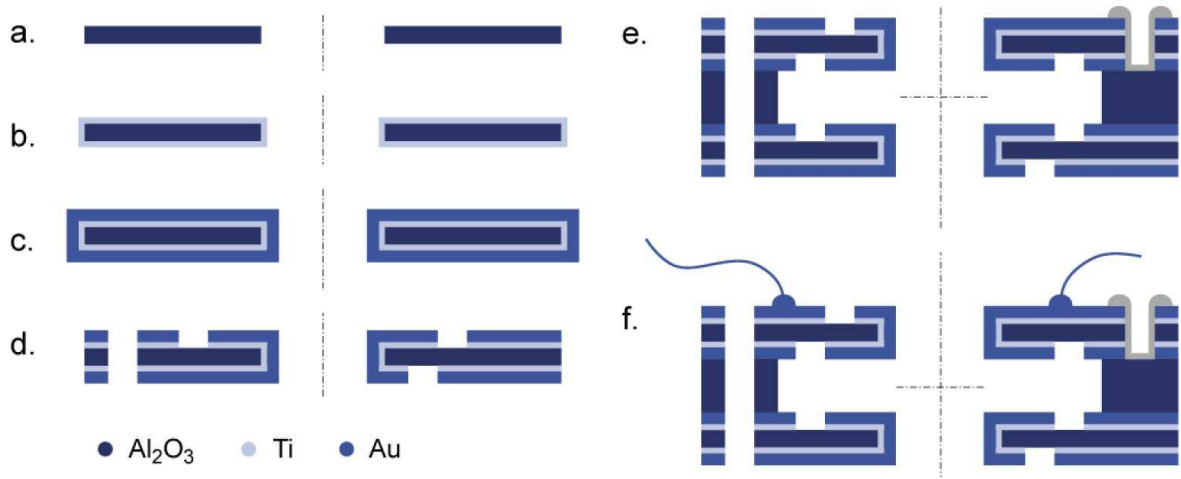}}
\caption{Fabrication process of the microtrap: processing of the Al$_2$O$_3$ wafer
includes laser machining (a), cleaning and coating (b)/(c) and laser cutting (d). Assembly
includes mounting (e) with adhesive (grey) and bonding (f).}
\label{konstruktion}
\end{figure}

The Ulm microtrap is based on a geometric three-layer stack design fabricated using gold coated
and laser structured Al$_2$O$_3$ wafers\footnote{Reinhardt Microtech AG, Wangs, Switzerland}. The
development process of the wafers starts with micro-machining of blank wafers (figure~\ref{konstruktion}a) using a femtosecond pulsed laser
source\footnote{Micreon GmbH, Hannover, Germany}.
In the storage zone the central slit width is h=500$\mu$m, while the transfer region
narrows the slit to g=250$\mu$m according to the width of the processing zone (figure~\ref{design}b). The
central slit in the storage/transfer zone and the processing zone is bounded by the DC electrode
fingers with a width of d=250$\mu$m and w=100$\mu$m separated with a spacing of
30$\mu$m. The constant length of the DC electrode fingers is 200$\mu$m.
The length of the RF notches with 60$\mu$m is optimized for mechanical stability and provides the axial
field symmetry with a 30$\mu$m spacing. Additional holes with a diameter of 240$\mu$m in the outer
region allows alignment and mounting.

The wafer metallization is done after a careful ultrasonic cleaning procedure
(acetone, isopropyl, piranha) and an oxygen plasma cleaning. The laser cutted blank
Al$_2$O$_3$ wafer is coated in an electron beam evaporator during a
continuous rotation with 50nm titanium and 500nm gold (figure~\ref{konstruktion}b and c). A declination angle
of 45$^{\circ}$ is required during the coating process for an overall continuous coating.
The surface roughness is better than 10nm, which was verified by atomic force microscopy.
 The conductor paths are manufactured by laser structuring of the wafer metallization (figure~\ref{konstruktion}d).
Finally, the wafer is laser diced. Bottom, center and top layer of the microtrap are
aligned and mounted with UHV compatible UV epoxy adhesive (figure~\ref{konstruktion}e).
The stack of three layers is mounted in a center holed UHV compatible leadless ceramic
84-pin chip carrier\footnote{Kyocera, type LCC8447001} with outer dimensions of 30mm
squared and an inner cavity size of 12mm. Electrical connections to the ceramic chip
carrier rely on ball bonding with 15$\mu$m diameter gold wire
(figure~\ref{konstruktion}e). Our design avoids dielectric blank Al$_2$O$_3$ areas,
unavoidable isolation lines between the electrode segments are shifted away
$>$200$\mu$m from the ions position by the extended electrode finger design. From an
electron microscopy analysis we checked that the DC electrode finger segments are
gold plated completely from all sides.

\begin{figure}[htp]
\resizebox{0.99\hsize}{!}{\includegraphics*{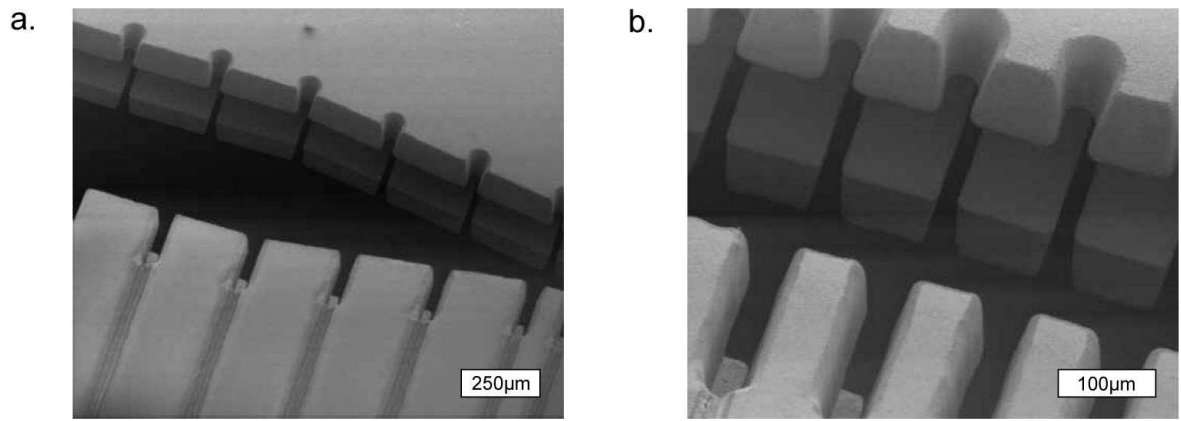}}
\caption{Electron microscope picture of the loading region (a) and narrower processing region (b).
The gold-coated finger structure and laser cuts defining electrodes are visible.}
\label{rem}
\end{figure}

The microtrap chip carrier is mounted on a low-pass filter printed circuit board (PCB),
that serves as electrical connector device of the 80-pin ribbon cables. The PCB is based
on the UHV compatible polyimide laminate Isola P97\footnote{Isola GmbH, D\"uren, Germany},
which is 200$\mu$m gold electroplated. The UHV compatible
Kapton ribbon cables are connected with ceramic D-type connectors in vacuum.
Low-pass filters (ceramic SMD type) with a cutoff close to 10MHz are soldered
close to the chip carrier (figure~\ref{design}a).

Decreasing electrode geometries of the processing zone compared to the storage zone of
the microtrap has some advantages with respect to quantum information experiments: A
smaller electrode design provides stronger confinement of the ions at the same voltage
levels. Stronger confinement leads to higher axial vibration frequencies,
which set the time scale for multi-qubit gate operations. For the details of the
trapping fields in axial and radial direction we use numerical simulations.

\subsection{Field simulations}
\begin{figure}[htp]
\resizebox{0.99\hsize}{!}{\includegraphics*{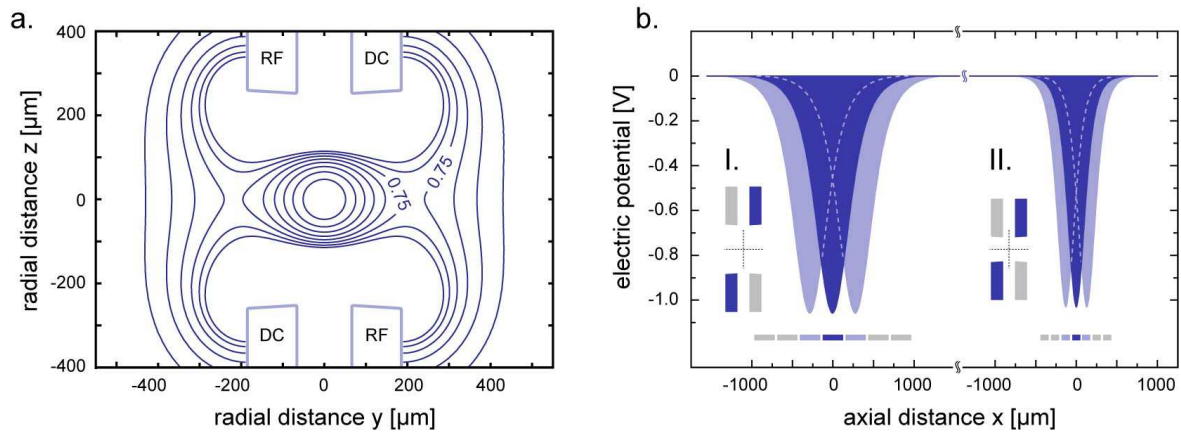}} 
\caption{Numerical
field simulation of the microtrap: (a) The pseudopotential cross section of the loading
region shows equi-pseudopotential lines at 0.125, 0.25, 0.375, 0.5, 0.625 and
0.75eV for a $^{40}Ca^+$ ion trapped at $\Omega=(2\pi)$~24.841MHz with $U=140$V.
The trap depth is 0.755eV. (b) The axial potential is plotted along the trap axis in the
loading (I) and processing (II) region. The individual electric potential of a single
electrode pair at -5V (other electrodes at 0V) is shown for three adjacent electrode
pairs.} \label{elfelder}
\end{figure}

The trap region of the microtrap is partitioned into a 9-segment loading and storage
section, a 3-segment transfer region and a 19-segment processing zone. Each electrode pair 
of a segment is voltage controlled separately, providing a full control for each
trap site individually and an effective micromotion compensation. Numerical simulations
give insight how to form potentials suitable to trap single ions or linear crystals
(figure~\ref{elfelder}). The loading zone is optimized such that a thermal beam of calcium
passes through, therefore the cross section of the storage zone is asymmetric with a
ratio 4:1. The more symmetric cross section of the processing zone with a ratio 2:1
leads to a strong confinement of the ions and higher radial and axial frequencies
\cite{SCHULZ2006}.

The 2D dynamical confinement in the radial cross section at the storage
region (figure~\ref{elfelder}a) is described numerically by the quadrupole potential
strength of the dynamical trap potential $\phi$. The lowest-order approximation
${\phi=c_2/2\,(y^2-z^2)\cdot U(t)}$ shows the geometric factor $c_2$ of the quadrupole
potential strength (yz cross section). The storage region is characterized by
$c_2=0.52\cdot 10^7\,m^{-2}$; the processing region shows a stronger confinement with
$c_2=1.99\cdot 10^7\,m^{-2}$. Under typical operating conditions, see
section~\ref{trapoperation} and with the trap drive frequency denoted by $\Omega/2\pi$,
the dimensionless stability parameter ${q=2eU/(m\Omega^2)\,c_2}$ results in ${q=0.14}$
for the storage and ${q=0.55}$ for the processing region. The calculated simplified
frequency of the secular motion ${\omega=\Omega\cdot q/2\sqrt{2}}$ in the storage
region is ${\omega=(2\pi)}$~1.26MHz and is reproduced experimentally with high
accuracy. The measurements presented here are implemented in the storage region, so the
processing region allows a tighter confinement of the ion in future experiments.

The axial potential along the trap axis is calculated in a numerical three-dimensional
electric potential simulation (figure~\ref{elfelder}b). Requirements for fast ion
transport are deep axial potentials with moderate control voltages and a large spatial
overlap of the axial potentials from adjacent electrode pairs \cite{SCHULZ2006}. The peak
width at half-height of the axial potentials are 500$\mu$m at the storage region
(250$\mu$m segment width) and 264$\mu$m at the loading region (100$\mu$m segment
width). For the wider segments of 250$\mu$m in the storage zone the trap allows a tight
confinement with an axial frequency of 1.20MHz with -5V applied only, which is
experimentally verified within 5\% accuracy via sideband spectroscopy.

\section{Experimental set-up}
\subsection{Trap operation}
\label{trapoperation} For the RF supply of the trap a RF synthesizer is used, followed
by a an amplifier to feed +26dBm ($\sim$400mW) power into a helical resonator matching
the 50$\Omega$ amplifier output impedance with the trap. The injected RF power is
monitored by an capacitive divider. Under typical operating conditions we reach
amplitudes of 280V$_{\rm pp}$ at a frequency of 24.841MHz. The DC electrodes are
supplied with voltages in the range of $\pm$10V. Here the ion is trapped in the middle of
the storage region with one electrode pair; the trap electrodes and the adjacent pairs
are biased for micromotion compensation. The trap is installed in a stainless
steel DN200CF vacuum chamber with a regular octagon DN63CF viewport symmetry. With a
pump system comprising an 75l ion pump and a titanium sublimation pump we operate the
experiment at a pressure of 10$^{-10}$mbar. Pairs of magnetic field coils in a
three-axis Helmholtz configuration are used for compensating the stray magnetic fields
and generating a quantization axis with a magnetic field of 0.3mT directed at
-45$^{\circ}$ with the trap axis x.

\begin{figure}[htp]
\resizebox{0.99\hsize}{!}{\includegraphics*{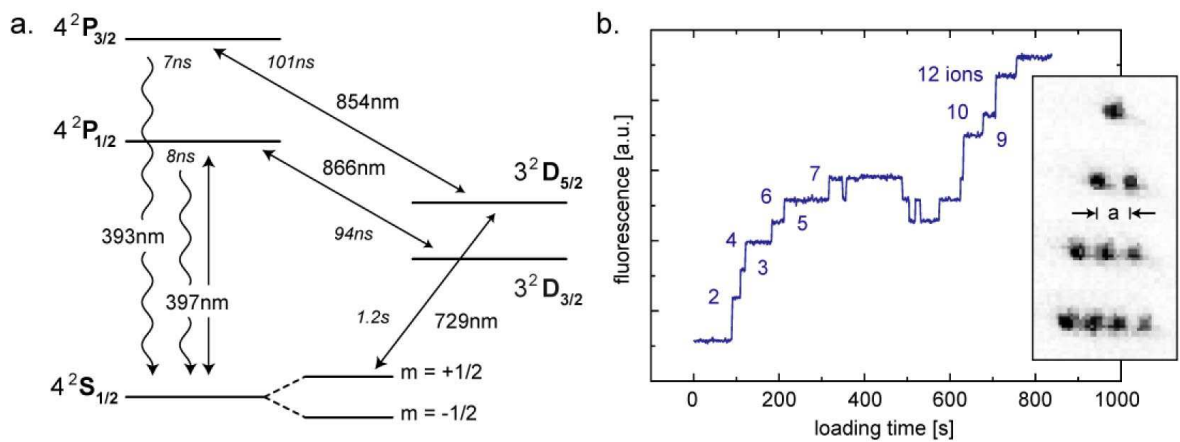}} \caption{(a)
Relevant levels, transition wavelengths and lifetimes in $^{40}Ca^+$. (b)~Fluorescence
near 397nm during loading observed with the EMCCD camera. The distance a of two ions
results in 7$\mu$m.} \label{levelscheme}
\end{figure}

\subsection{Laser configuration}
Single $^{40}$Ca$^+$ ions stored in the trap are generated by photoionization of a
neutral calcium atomic beam from a resistively heated oven. The two step process is
driven by laser light near 423nm and 375nm \cite{GULDE2001}. Figure~\ref{levelscheme}a shows
the level scheme of $^{40}$Ca$^+$: The Dipole transitions
4$^2$S$_{1/2}\rightarrow$~4$^2$P$_{1/2}$ at 397nm allows Doppler cooling of the
ions, the 3$^2$D$_{3/2}\rightarrow$~4$^2$P$_{1/2}$ near 866nm and the transition
3$^2$D$_{5/2}\rightarrow$~4$^2$P$_{3/2}$ near 854nm the depletion of the metastable
D-levels. The quadrupole transition near 729nm from the 4$^2$S$_{1/2}$ ground state to
the metastable 3$^2$D$_{5/2}$ level is employed for sideband
spectroscopy, sideband cooling and coherent quantum dynamics.

All transitions of the $^{40}$Ca$^+$ ion are driven by grating stabilized diode lasers.
The lasers at 397nm, 866nm and 854nm are locked to external Zerodur Fabry-Perot
cavities (finesse F=250) for frequency stabilisation using the Pound-Drever-Hall (PDH)
technique. The laser at 729nm (laser diode followed by a tapered amplifier) is PDH-locked 
to a cavity with finesse $F \geq 50000$ reaching a sub-kilohertz linewidth; the cavity 
is fabricated from ultra low expansion glass (ULE) material. Lasers at
866nm, 854nm and 729nm can be switched off using acousto-optical modulators in
double-pass configuration, also being employed for tuning the laser frequency near
729nm. Laser beams at 397nm, 866nm and 854nm are within the xy plane (see
figure\ref{elfelder}) and intersect the trap at $+$45$^{\circ}$ with respect to the trap
axis x. A separate $\sigma+$ beam is deviated from the laser near 397nm for optical
pumping. Also this beam lies in the xy plane but intersects the trap axis under
-45$^{\circ}$, parallel to the magnetic field direction. The beam near 729nm, in the xy
plane, is focused to a waist size of 15$\mu$m and intersects the trap under
45$^{\circ}$ perpendicular to the magnetic field axis. The Lamb-Dicke factor of
excitation depends on the projection of the $\overrightarrow{k}_{729}$ vector on the
trap axis as well as the trap frequency $\omega_{ax}=(2\pi)$~1.1MHz and mass m of the
ion like $\eta_{729}=|\overrightarrow{k}_{729}|\:{\rm cos}\: \theta \sqrt{h/2m
\omega_{\rm ax}}$ and we calculate $\eta_{729}$=0.065. The polarization at 729nm is
chosen such that only $\Delta$m=2 transitions are allowed, e.g. from $S_{1/2}$, m=+1/2
to $D_{5/2}$, m=+5/2 and m=-3/2.

\subsection{Doppler cooling, fluorescence detection and cold ion crystals}
The ion fluorescence is imaged by a custom lens system\footnote{Sill Optics,
Wendelstein, Germany} (focal length 66.8mm, numeric aperture 0.27)
 on an EMCCD camera\footnote{iXon DV860-BI, Andor
Technology, Belfast, Northern Ireland} (efficiency of 50\%) and a photomultiplier tube
(PMT) of quantum efficiency $\sim$27\% for detecting ions and reading out the quantum
state. The magnification of the imaging branch is roughly 20. The fluorescent light is
collected from a solid angle of $\sim$2.5\%. It is distributed at a ratio of 80:20
between PMT and camera by a beam splitter and filtered by a band pass filter to
suppress background photons. For trapping and cooling, 1mW of 866nm light is focussed
to a spot size of 30$\mu$m. The power of the laser near 397nm is focussed to a spot size of 20$\mu$m
and can be switched between zero, a reduced power level of 30$\mu$W (Doppler cooling)
and full power of about 300$\mu$W (detection). With the laser slightly red detuned
and proper micromotion compensation, a single ion count rate of 16kHz at a background
of 4kHz can be achieved with the PMT. If an ion is shelved in the long lived D$_{5/2}$
level, no light is scattered. With a detection time of 5ms, we are able to discriminate
the qubit states with an error probability of $3\cdot 10^{-3}$ \cite{ROOSPHD}.
Figure~\ref{levelscheme}b shows linear ion crystals trapped and observed in the
microtrap.

\section{Quantum jump spectroscopy}
Quantum jump spectroscopy has been used to determine the frequency
of clock transitions with an accuracy of about 7 parts in 10$^{-17}$
\cite{ROSENBAND2007,OSKAY2006,MARGOLIS2004,SCHNEIDER2005}. A narrow
dipole forbidden transition is driven and subsequently the
excitation to the metastable level is tested by exposure to resonant
radiation on a dipole allowed transition.

\subsection{Sideband spectroscopy on the S$_{1/2}$ to D$_{5/2}$ transition}
In the experiments presented here, we perform spectroscopy on the
S$_{1/2}$ to D$_{5/2}$ transition (figure~\ref{seitenband}). With a lifetime of 1.2s, the
spectroscopic resolution is limited by the laser pulse duration, the
Rabi frequency during the excitation and the frequency stability of
the laser source. The harmonic motion of the ion in the trap can be
investigated spectroscopically at the level of single vibrational
quanta. The electronic and vibrational state is manipulated
coherently with the following sequence:

\begin{figure}[htp]
\resizebox{0.7\hsize}{!}{\includegraphics*{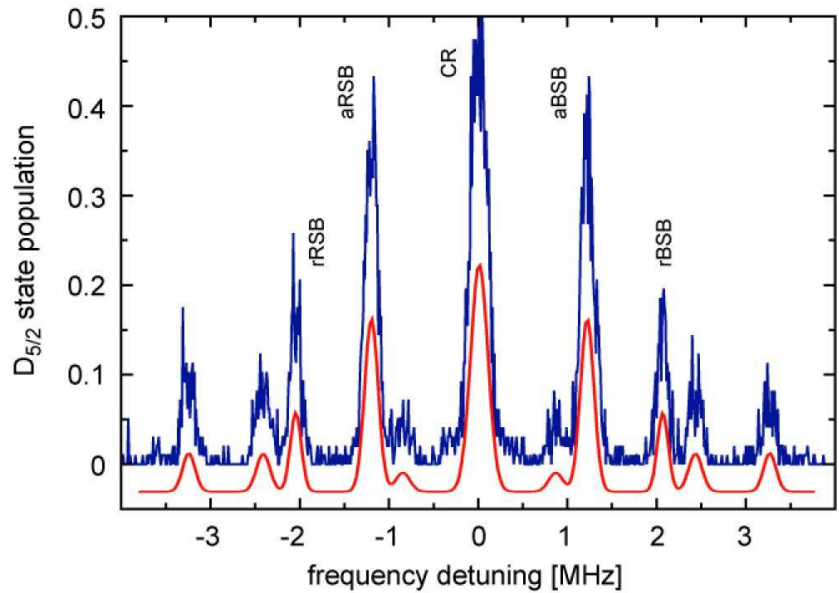}} \caption{Quantum
jump spectroscopy with a single ion: the laser frequency near 729nm is scanned over the
carrier transition and the vibrational sidebands. For the data shown here, we have used
a laser linewidth of $\sim$200kHz. The fit to the data (shifted and downsized
for clarity) determines the axial and radial trap frequencies,
$\omega_{\rm ax}$=(2$\pi$)~1.2MHz and $\omega_{\rm rad}$=(2$\pi$)~2MHz.} \label{seitenband}
\end{figure}

\begin{enumerate}
\label{sequence}
    \item Doppler cooling: the laser near 397nm is red detuned from the S$_{1/2}$ to P$_{1/2}$
    transition to $\lesssim$1/2 of the maximum fluorescence rate. This corresponds to a setting of about
    $\Gamma/2$ where $\Gamma$=(2$\pi$)~22.3MHz is the natural linewidth of the dipole transition.
    The beam is attenuated in order to avoid
    saturation. The laser frequency near 866nm is tuned for maximum fluorescence. Additionally,
    resonant laser light depopulates the metastable D$_{5/2}$ level. Typically we apply
    Doppler cooling for 5ms. While the theoretical cooling limit results in a mean phonon number of
    $\overline{n} \hspace{2mm} \hbar \omega_{\rm trap} = \Gamma/2$, with $\overline{n}\sim$12, typically we
    observe a slightly higher $\overline{n}$ between 12 and 25.
    \item Optical pumping: with a typically 5$\mu$s short pulse of $\sigma+$ polarized light near 397nm we
    pump the ion into the S$_{1/2}$, m=+1/2.
    \item Sideband cooling (optionally): we tune the laser light near 729nm such that the red secular
    sideband of the S$_{1/2}$, m=+1/2 to D$_{5/2}$, m=+5/2 is excited. The D$_{5/2}$ state is quenched by resonant laser light near  854nm to the P$_{3/2}$ level which quickly decays to S$_{1/2}$ closing the cooling
    cycle. Short pulses of optical pumping as in step (ii) are inserted into, and  also conclude the sideband cooling.
    \item Spectroscopy pulse: all laser sources, except the laser light near 729nm, are switched off.
    We excite the S$_{1/2}$, m=+1/2 to D$_{5/2}$
    transition with pulses of well-defined frequency, duration and intensity.
    \item State detection: the lasers at 397nm and 866nm are switched back on. The power of the 397nm laser is at maximum
    such that a maximum count rate is obtained. The
    PMT counts photons of the ions fluorescence. If the ion had been excited on the quadrupole
    transition to the D$_{5/2}$ level, no fluorescence photons will be observed.
\end{enumerate}

The entire sequence is repeated typically 50 to 500 times depending
on the requirements, and the average excitation to the D$_{5/2}$
level is recorded. For $N$~measurements and an excitation
probability $p$, the projection noise error is given by
$\sqrt{p(1-p)/N}$. Then e.g. the
laser frequency near 729nm is varied, see figure~\ref{seitenband}.
Here, the ion is excited without sideband
cooling step (iii) such that the radial and axial red and blue
motional sidebands show equal strength. These resonances are denoted
with rRSB to rBSB. Additional resonances show up at a laser detuning
for a second sideband excitation $\pm 2\omega_{\rm ax}$ and at
difference frequencies of the sidebands~$\omega_{\rm rad}-\omega_{\rm ax}$.

\begin{figure}[htp]
\resizebox{0.7\hsize}{!}{\includegraphics*{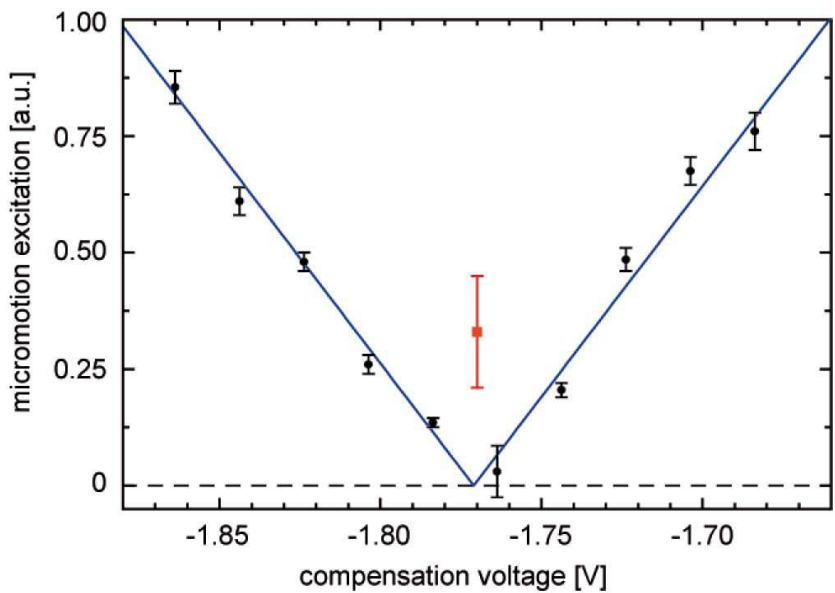}}
\caption{Excitation on the micromotional sideband for compensation: the amplitude of
the excitation line is plotted as the compensation voltage is varied. For comparison,
the excitation amplitude of the carrier with a laser power reduced by a factor of 10
(indicated by the red square data point).} \label{kompensation}
\end{figure}

\subsection{Micromotion compensation}
If an ion is displaced from the node of the RF electric field either by asymmetries or
by patch charges, it will oscillate at the trap drive frequency $\Omega/2\pi$.
Consequently, Doppler cooling and fluorescence detection will be affected. However,
micromotion is compensated by applying a balanced voltage difference to the particular
segments of an electrode pair such that the ion is moved into the RF trap center.
Various methods of measuring the micromotion have been studied \cite{BERKELAND1998};
yet another method uses sideband spectroscopy: we excite the sideband at $\omega_{\rm
carrier}+\Omega$  and compare the sideband excitation with that on the carrier
\cite{ROOSPHD}. The Rabi frequency $\Omega_1$ on the micromotion sideband and that one
on the carrier $\Omega_0$ holds $\Omega_1$/$\Omega_0$ = J$_1$($\beta$)/$J_0$($\beta$)
$\approx \beta$/2 for $\beta \ll 1$, with $\beta$ denoting the index of modulation and
J$_n$ indicating the Bessel functions of the n-th order. As the excitation to the
D$_{5/2}$ state is proportional to $\Omega^2$ for low saturation, we can measure the
ions micromotion directly, see figure~\ref{kompensation}. We find the minimum excitation
close to -1.77V. At this minimum, thus for optimum compensation voltage, we reduce the
ratio of excitation strength to zero with an error of $\pm$0.03. This measured value
corresponds to the ratio of squared Bessel functions J$^2_1$($\beta$)/J$^2_0$($\beta$)
with modulation index of $0.0\pm0.17$, which is due to a residual micromotion
oscillation amplitude x$_{\rm min}=\beta_{\rm min}(\lambda/2) =
(0.0\pm0.17)(729\textrm{nm}/2)$ of (0$\pm$130)nm. The optimal compensation voltage
changes by less than 0.5\% from day to day.

In conclusion, the micromotion can be nulled by a properly set compensation voltage
and, due to the shielding effect of the gold-coated finger-shaped electrodes, stray
electric fields from isolating parts of the microtrap do not show up with large
and fluctuating contributions.

\begin{figure}[htp]
\resizebox{0.7\hsize}{!}{\includegraphics*{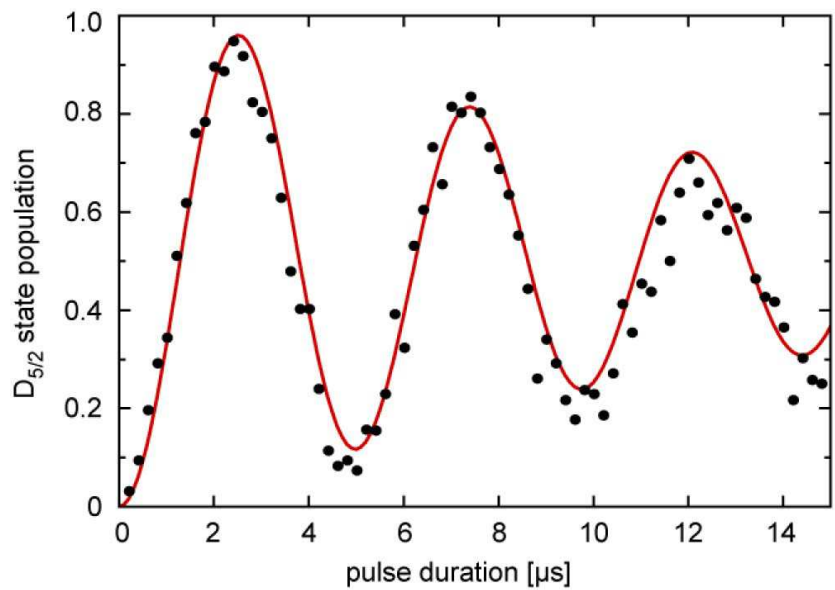}} \caption{Rabi
oscillation with a single ion on the carrier transition. The data are taken with a
laser power of $\sim$60mW. The model curve assumes the dephasing of the oscillation due
to a thermal distribution with a mean phonon number of $\overline{n}\sim$12.} \label{rabi}
\end{figure}

\subsection{Coherent single ion dynamics}
The coherent ion-light interaction leads to Rabi oscillations
between the S$_{1/2}$ ground state and the excited metastable
D$_{5/2}$ state when the duration of the interaction is varied. If
the single ions internal states are used to store qubit
information, a $\pi$-pulse will flip the qubit between the two logic
states.

Here, we present Rabi oscillations on the carrier transition for a single ion, see
figure~\ref{rabi}, which has been Doppler-cooled. The data shows a 95\% efficiency for the
$\pi$-pulse at 2.5$\mu$s, corresponding to a Rabi frequency of $\Omega_0=(2\pi)$
200kHz. The decay of contrast is due to the fact that we observe an incoherent
superposition of Rabi oscillations with a different frequency for each thermally
occupied Fock state: $\Omega_{n,n}\propto 1-\eta^2_{729}n$. If we model the data with
$P_{D}(t)=\sum{p_n(\overline{n}) {\rm sin}^2(\Omega_{n,n}(t))}$ we find agreement for a
thermal distribution p$_n$ with $\overline{n}\sim$12 (figure~\ref{rabi}).

\begin{figure}[t]
\resizebox{0.99\hsize}{!}{\includegraphics*{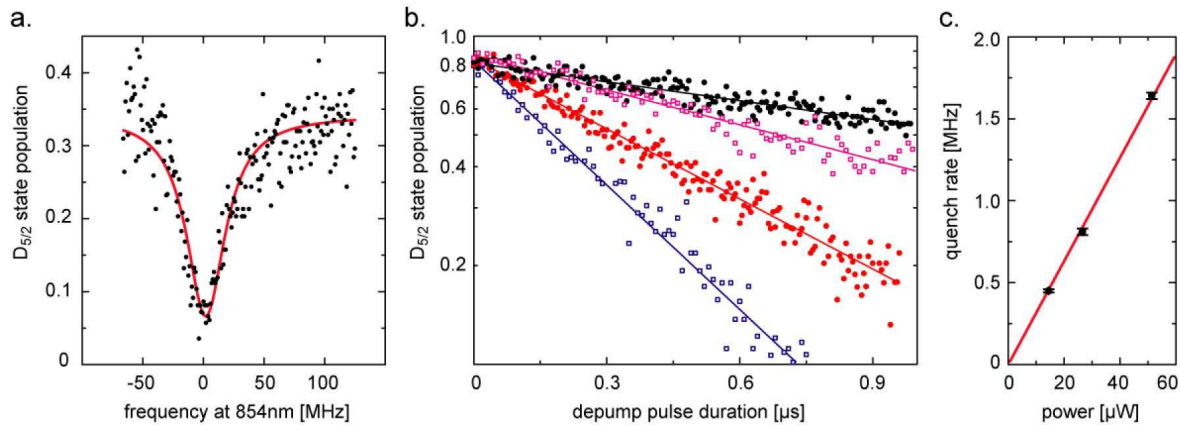}} \caption{(a) Depletion
of the D$_{5/2}$ state by a 3$\mu$s pulse near 854nm. The observed dip corresponds the
resonance line and allows for tuning the laser to resonance. We model the data by a
37MHz power-broadened Lorentzian. (b) After a $\pi$-pulse on the carrier transition at
729nm, the depletion pulse length to the P$_{3/2}$ state is scanned. The exponential
decay fit determines $\gamma_{\rm eff}$, here plotted with four different laser powers at
854~nm. (c) The dependence of $\gamma_{\rm eff}$ is plotted as a function of the 854nm laser
power, a linear fit yields 31.6(5)kHz/$\mu$W. \label{las854}}
\end{figure}

\subsection{Sideband cooling}
For Doppler cooling, the temperature limit is given by the natural
linewidth of the dipole transition. For quantum logic operations,
however, a mean phonon number below this limit is often required.
Therefore we apply sideband cooling on the narrow S$_{1/2}$ to
D$_{5/2}$ transition \cite{MARZOLI}. Laser radiation near 729nm on
the red sideband of the transition excites from S$_{1/2}$, m=+1/2 to
D$_{5/2}$, m=+5/2. The effective width of this cooling transition
can be increased by applying resonant laser light near 854nm. This
mixes the D$_{5/2}$ state with the P$_{3/2}$ state which rapidly
decays to the S$_{1/2}$, m=+1/2. The cooling rate is modified
accordingly, and the effective width of the D$_{5/2}$ level sets an
upper limit for the rate of cooling cycles.

For the introduction of laser cooling of trapped particles we apply
the semiclassical theory \cite{STENHOLM1986}, determine all laser
cooling parameters experimentally and compare the cooling result
with the theoretical expectation.

The theoretical limit of sideband cooling is given by the ratio of
laser cooling $\Gamma_{\rm cool}$ and the heating rates
$\Gamma_{\rm heat}=\Gamma_{\rm laser-heat}+\Gamma_{\rm trap}$, yielding
$\overline{n}=\Gamma_{\rm heat}/(\Gamma_{\rm cool}-\Gamma_{\rm heat})$ as the steady
state average phonon number. Heating by laser processes is due to
either an off-resonant excitation on the carrier transition with
subsequent decay on the blue sideband or an off-resonant blue
sideband excitation followed by a decay on the carrier
\cite{STENHOLM1986}. A calculation of the detailed balance of phonon
states leads to \begin{eqnarray}
  \overline{n}=\left(\frac{\eta^2_{\rm spont}}{\eta^2_{729}}+\frac{1}{4}\right)\hspace{1mm}
  \frac{\gamma_{\rm eff}^2}{4\omega_{\rm ax}}
  \label{cool1}
\end{eqnarray}
if trap heating is excluded. Here, the parameter $\eta_{\rm spont}
=|\overrightarrow{k}_{395}| \sqrt{h/2m \omega_{\rm ax}}$ results in 0.17, for the
axial trap frequency~$\omega_{\rm ax}$ =$(2\pi)$ 1.1MHz, as it takes into account the
recoil if the ion decays spontaneously from the P$_{3/2}$ level to the S$_{1/2}$ ground
state. Due to the laser recoil on the S$_{1/2}$ to D$_{5/2}$ excitation we get
$\eta_{729}$=0.065. In equation~\ref{cool1} only the cooling rate, but not the cooling limit
depends on the intensity of the laser near 729nm. $\gamma_{\rm eff}$ denotes the
effective linewidth from quenching the D$_{5/2}$ state to the P$_{3/2}$ and is adjusted
by the laser power at 854nm.

\begin{figure}[t]
\resizebox{0.99\hsize}{!}{\includegraphics*{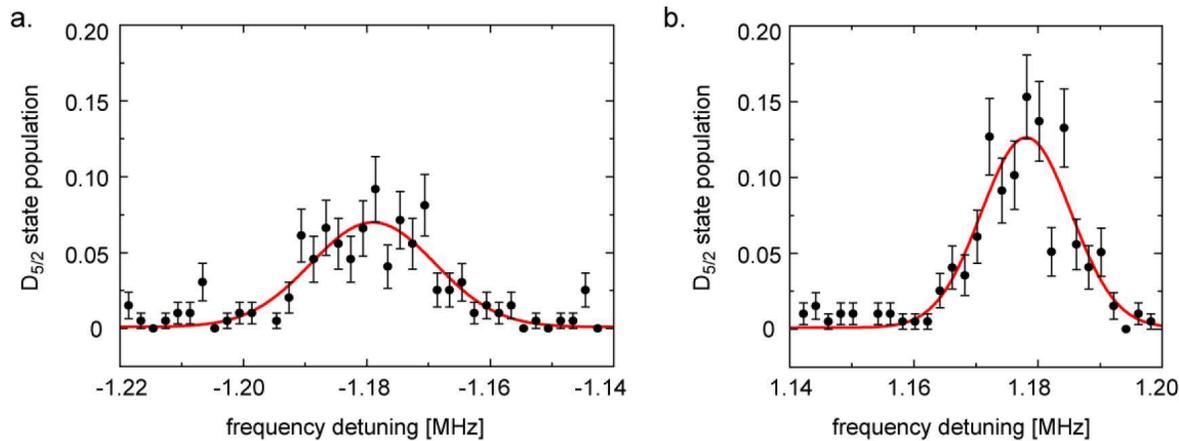}} \caption{(a) Sideband
cooling of the axial vibration of a single ion. The imbalance of blue and red sideband
intensities P$_{\rm blue}$ and P$_{\rm red}$ leads to an estimation of a mean phonon number $\overline{n}$
 of 1.2(3). The spectrum has been
measured with a 200$\mu$s delay after the sideband cooling. (b)~The trap heating is
determined from a linear fit to the data, see figure~\ref{heizrate}.} \label{sbcool}
\end{figure}

The average phonon number~$\overline{n}$
increases with the intensity of the laser near 854nm; thus one has
a tradeoff between cooling rate and minimum temperature. However,
even for $\gamma_{\rm eff} \simeq$ 90kHz, equation~\ref{cool1} predicts an
almost perfect ground state of vibration with only
$\overline{n}\leq$0.01. The situation is more complicated if we take
trap heating into account: the steady state photon number results
from the balance

 \begin{eqnarray}
  \overline{n}=\frac{\Gamma_{\rm laser-heat}+\Gamma_{\rm trap}}
  {\Gamma_{\rm laser-cool}-\Gamma_{\rm laser-heat}-\Gamma_{\rm trap}}
  \simeq \frac{\Gamma_{\rm trap}}
  {\Gamma_{\rm laser-cool}-\Gamma_{\rm trap}}
\end{eqnarray}
if the laser induced heating is small compared to the trap heating
rate. We find that the thermal mean phonon number becomes
\begin{eqnarray}\label{cool2}
\overline{n}=\frac{\Gamma_{\rm trap}}{(\eta_{729}\Omega_{0}/\gamma_{\rm eff})^2 \hspace{2mm}
\gamma_{\rm eff} -\Gamma_{\rm trap}},
\end{eqnarray}
for the case where the net cooling rate $W=(\eta_{\rm 729}\Omega_{\rm 0})^2/\gamma_{\rm eff} -
\Gamma_{\rm trap}$ is positive. In contrast to equation~\ref{cool1}, the cooling
limit now depends on the intensity of the laser near 729nm driving
the red sideband of the quadrupole transition. We consider here the
case where the sideband excitation is incoherent, with $\eta_{\rm 729}\Omega_{\rm 0}
\leq \gamma_{\rm eff}$.

In order to compare the above cooling theory to the experiment, we
need to determine $\gamma_{\rm eff}$, $\Omega_{0}$ and $\Gamma_{\rm trap}$
in independent experimental measurement sequences and compare the
theoretical prediction in equation~\ref{cool2} with the experimental
outcome on $\overline{n}$.

$\gamma_{\rm eff}$ is determined by controlled depletion of the
metastable state: After Doppler cooling
and optical pumping, we apply a 1$\mu$s laser pulse on the carrier
transition such that the ion is transferred into the D$_{5/2}$ state. A pulse
of laser light resonant to the D$_{5/2}$ to P$_{3/2}$ transition is
then applied and finally the remaining D$_{5/2}$ population is
determined. From the exponential decay, see figure~\ref{las854}b,
we determine the effective cooling width that is as expected a
linear function of the laser power at 854nm, with
$\gamma_{\rm eff}[$kHz$]=31.6(5)$P$_{854}[\mu W]$.

The Rabi frequency on the quadrupole transition is revealed from measurements such as
in figure~\ref{rabi}. With the maximum available laser power of 60mW we reach
{$\Omega_0 \simeq (2\pi)$~200kHz}. Correspondingly, the maximum possible sideband excitation
with this laser power is {$\eta_{729}\Omega_0$ =(2$\pi$)~13kHz}.

\begin{figure}[t]
\resizebox{0.6\hsize}{!}{\includegraphics*{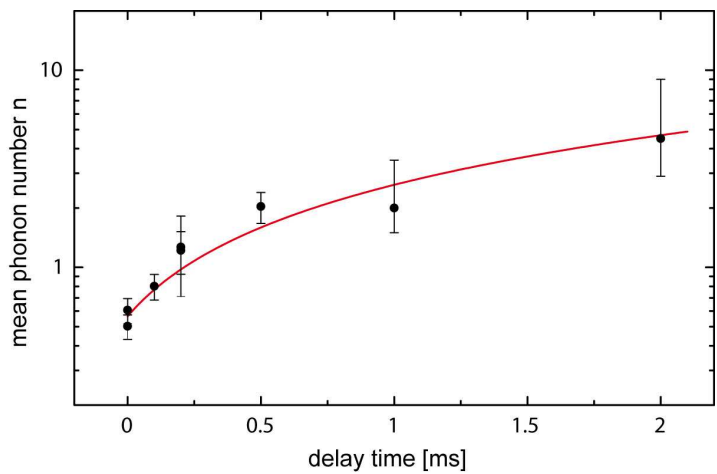}} \caption{Heating rate
measurement of the axial vibrational mode of a single ion as measured from the
temperature increase after a delay time. The values of $\rm d\overline{n}/dt$ = 2.1(3)
per ms and the minimum phonon number of 0.56(5) are determined from a linear fit to the
data.} \label{heizrate}
\end{figure}

\subsection{Heating rate measurement}
The mean phonon number of $\overline{n}$ is determined from spectra such as in
figure~\ref{sbcool} showing the red and blue sideband after sideband cooling. We deduce
from the ratio of the sideband intensities A=P$_{\rm red}$/P$_{\rm blue}$ by means of
the asymmetry of the excitation (figure~\ref{sbcool}) the mean phonon number
$\overline{n}={\rm A/(1-A)}$, here $\overline{n}$=0.56(5). To obtain the trap heating
rate, we insert a variable waiting time interval between steps (iii) and (iv) in the
experimental sequence, see section~\ref{sequence} \cite{ROOS1999}. The data plotted in
figure~\ref{heizrate} yields a trap heating of 2.1(3) per ms. As this heating rate is
attributed to the contamination of trap electrodes due to the loading process from the
thermal beam of neutral calcium, we expect a lower mean phonon number $\overline{n}$ in
the processing region, which is spatially separated from the loading zone.

Note that quantum gate operations and the transport of ions in the trap are about 10 to
100 times faster than the measured trap heating. In future we will investigate the
influence of the gold coating (different surface qualities and thickness) on the trap
heating.

\section{Conclusion and outlook}
We have outlined the design, the fabrication and the first characterization of a novel
multi-segmented microchip ion trap, and we have show that this trap is suited for
scalable quantum logic. In some detail, the sideband cooling on the quadrupole
transition has been investigated. In future, we will explore Raman transitions between
Zeeman ground (qubit) states S$_{1/2}$, m=+1/2 and m=-1/2 for the cooling, for coherent
qubit rotations and for two-qubit gate operations. Additionally, we will study the
transport of quantum information between the processor and the memory section
benefiting from the large number of trap segments. Further development will include the
integration of micro-optical elements.

\vspace{1cm}

\textbf{Acknowledgements:} We thank Frank Korte (Micreon GmbH) for his expertise in
laser cutting, H.~Roscher for the cooperation in the cleanroom facility, K.~Singer and
R.~Reichle for their contributions in an earlier stage of the experiment, K.~Singer for his
important help with the experimental control software, and J.~Eschner for discussions.
We acknowledge financial support by the German science foundation DFG within the
SFB/TRR-21 and by the European commission within MICROTRAP (Contract No.~517675) and
EMALI (Contract No. MRTN-CT-2006-035369).

\vspace{1cm}

\bibliographystyle{unsrt}
\bibliography{lit}

\end{document}